\documentclass[prd,aps,twocolumn,a4paper,floatfix,showpacs,nofootinbib]{revtex4-1}

\usepackage[utf8]{inputenc}  
\usepackage[T1]{fontenc}
\usepackage{graphicx,psfrag}
\graphicspath{{figures/}}
\usepackage{mathrsfs}
\usepackage{amsmath,amsfonts,amssymb}
\usepackage{multirow,enumerate}
\usepackage{comment,hyperref}
\usepackage{color}
\usepackage{acronym}
\usepackage{xspace}
\usepackage[normalem]{ulem}
\usepackage{mathtools}
\usepackage{subfigure}
\usepackage{makecell}
\usepackage{appendix}
\usepackage{diagbox}
\usepackage{lineno}

\newacro{BH}{black hole}
\newacro{NS}{neutron star}
\newacro{PN}{Post-Newtonian}
\newacro{BBH}{binary black hole}
\newacro{BNS}{binary neutron star}
\newacro{EOB}{effective-one-body}
\newacro{NR}{numerical relativity}
\newacro{GW}{gravitational wave}
\newacro{EOS}{equation-of-state}

\newcommand{\be}{\begin{equation}}
\newcommand{\ee}{\end{equation}}
\newcommand{\bea}{\begin{eqnarray}}
\newcommand{\eea}{\end{eqnarray}}
\newcommand{\bel}{\begin{align}}
\newcommand{\eel}{\end{align}}

\def\nn{\nonumber}

\def\GMc2{{\rm G M_{\odot} c^{-2}}}

\def\SEOBNRv4T{\texttt{SEOBNRv4T}\xspace}

\usepackage{color}
\definecolor{cyan}{rgb}{0,0.9,0.9}
\definecolor{orange}{rgb}{0.9,0.5,0}
\definecolor{magenta}{rgb}{1,0,1}
\definecolor{purple}{rgb}{0.8,0.4,0.8}
\definecolor{gray}{rgb}{0.5,0.5,0.5}
\definecolor{mygreen}{rgb}{0.1,0.8,0.1}
\definecolor{darkblue}{rgb}{0.0,0.0,0.6}

\definecolor{mangotango}{rgb}{1.0, 0.51, 0.26}

\begin{document}

\title{Determining the equation of state of neutron stars with Einstein Telescope using 
tidal effects and r-mode excitations from a population of binary inspirals} 

\author{Pawan Kumar Gupta$^{1,2}$}
\author{Anna Puecher$^{1,2}$}
\author{Peter~T.H.~Pang$^{1,2}$}
\author{Justin Janquart$^{1,2}$}
\author{Gideon Koekoek$^{1,3}$}
\author{Chris Van Den Broeck$^{1,2}$}

\affiliation{${}^1$Nikhef -- National Institute for Subatomic Physics, 
Science Park 105, 1098 XG Amsterdam, The Netherlands}
\affiliation{${}^2$Institute for Gravitational and Subatomic Physics (GRASP), 
Utrecht University, Princetonplein 1, 3584 CC Utrecht, The Netherlands}
\affiliation{${}^3$Department of Gravitational Waves and Fundamental Physics, 
Maastricht University, P.O.~Box 616, 6200 MD Maastricht, The Netherlands}

\date{\today}

\begin{abstract}
Third-generation gravitational wave (GW) observatories such as Einstein Telescope (ET) and Cosmic 
Explorer (CE) will be ideal instruments to probe the structure of neutron stars through 
the GWs they emit when undergoing binary coalescence. In this work we make 
predictions about how well ET in particular will enable us to reconstruct 
the neutron star equation of state through observations of tens of binary neutron star
coalescences with signal-to-noise ratios in the hundreds. We restrict ourselves to  
information that can be extracted from the inspiral, which includes tidal effects and possibly 
r-mode resonances. In treating the latter we go beyond the Newtonian approximation, introducing 
and utilizing new universal relations. We find that the ability to observe resonant r-modes would have a 
noticeable impact on neutron star equation of state measurements with ET.
\end{abstract}

\maketitle

\section{Introduction}
\label{sec:intro} 

The detection with Advanced LIGO \cite{TheLIGOScientific:2014jea} and Advanced Virgo
\cite{TheVirgo:2014hva} of gravitational waves (GWs) from the binary neutron star (BNS) coalescences
GW170817 \cite{LIGOScientific:2017vwq} and GW190425 \cite{Abbott:2020uma}, together with 
electromagnetic observations \cite{DES:2017kbs,Cowperthwaite:2017dyu,LIGOScientific:2017zic}
has already had a significant impact on our insight into the structure of neutron 
stars; for a recent review, see e.g.~\cite{Dietrich:2020eud}. Even so, the neutron star equation of 
state (EOS) remains poorly constrained. This is expected to change with the advent of 
third-generation GW detectors such as Einstein Telescope (ET) \cite{Punturo:2010zza,Hild:2010id}
and Cosmic Explorer (CE) \cite{LIGOScientific:2016wof,Reitze:2019iox}, which are likely 
to see hundreds of thousands of binary neutron star coalescences, of which hundreds may 
have signal-to-noise ratios (SNRs) in excess of 100; see \cite{Samajdar:2021egv} for recent
estimates. Whereas with current detectors we only have access to the inspiral signal, ET and CE 
will also probe the post-merger \cite{Maggiore:2019uih,Evans:2021gyd}. 

In this paper we will quantify to what extent the equation of state -- essentially pressure 
versus density, $p(\rho)$ -- of neutron stars can be probed with third-generation GW observatories, based 
on inspiral physics alone. The main EOS-related effect that enters the inspiral signal 
is that of tides on the neutron stars \cite{Flanagan:2007ix,Hinderer:2007mb}. These can already be 
investigated with second-generation detectors \cite{DelPozzo:2013ala,Agathos:2015uaa,Lackey:2014fwa}, 
as has indeed been done with GW170817 
\cite{LIGOScientific:2017vwq,LIGOScientific:2018cki,LIGOScientific:2019eut}. 
Tidal effects are the most noticeable at high frequencies. However, given sufficiently high SNR, it may 
also be possible to see resonant r-modes \cite{Flanagan:2006sb}, gravitomagnetic excitations
of Rossby modes that are induced when the monotonically increasing GW frequency reaches  
an associated resonance frequency of one of the neutron stars. Assuming slowly rotating 
neutron stars, these will mainly manifest themselves at lower frequencies, i.e.~tens of Hz. 
Though the imprints of resonant r-modes on GW signals will generally be weak, they can be within 
reach of third-generation observatories \cite{Poisson:2020eki}. Moreover, as 
recently indicated by Ma et al.~\cite{Ma:2020oni}, EOS measurements may gain  
from the ability to access r-mode information in addition to tidal effects. In this study we will focus 
on ET, in part because it is predicted to have excellent low frequency sensitivity.

In \cite{Ma:2020oni}, Fisher matrix estimates were made to assess the measurability 
of tidal deformabilities and other parameters related to neutron stars in binary inspiral, 
with or without resonant r-modes. For the purpose of initial estimates, in the latter paper 
the r-modes were largely treated in the Newtonian approximation, but it is known that 
relativistic corrections on both the resonance frequency and the size of the induced 
GW phase shift can be sizeable; see e.g.~\cite{Idrisy:2014qca,Gupta:2020lnv,JimenezForteza:2018rwr}. 
Here we do take such corrections into account, and we introduce and use new universal relations
for both. Moreover, we want to know what is the impact of having a sizeable number of 
high-SNR BNS signals at one's disposal. Ideally we would want to perform 
Bayesian parameter estimation simulations, but limited computational resources prevent us from 
doing so. As a compromise, we will still use the Fisher matrix to obtain multivariate
Gaussian approximations for the likelihoods of individual sources, but we sample from those
to obtain estimates for EOS measurements with input from multiple detected sources. 
Recently, catalogs of binary black hole and binary neutron star detections by ET and CE 
were simulated \cite{Samajdar:2021egv} based on predictions for merger rate as a function 
of redshift; we will draw from this work to obtain a representative sample for the 
20 loudest BNS signals that ET is likely to detect based on its predicted
sensitivity curve. Based on these we will arrive at estimates for how accurately one
will be able to reconstruct $p(\rho)$ with observations by ET, with or without r-modes. 

The rest of this paper is structured as follows. In Sec.~\ref{sec:rmodes} we describe our
treatment of resonant r-modes in terms of universal relations. In Sec.~\ref{sec:setup}
we explain the setup of our analyses. Results are given in Sec.~\ref{sec:results}. Finally, 
an overview and conclusions are presented in Sec.~\ref{sec:conclusions}. Unless stated 
otherwise, throughout this paper we set $G = c = 1$.

\section{Resonant r-modes: Frequencies, phase shifts, and universal relations}
\label{sec:rmodes}


As two neutron stars spiral towards each other, the GW frequency 
increases monotonically, and at one or more points in time it can become equal to 
an internal resonance frequency of one of the neutron stars. The resulting excitation
takes away part of the orbital energy, which speeds up the orbital motion; this in turn
gets imprinted upon the phasing of the GW signal. Assuming the two neutron star 
each undergo near-instantaneous resonances at respective frequencies $f_0^{(1,2)}$,
the resulting change in the frequency-domain GW phase with respect to the point
particle case can be modeled as 
\cite{Flanagan:2006sb,Flanagan:2007ix}
\begin{eqnarray}
\Psi_r 
&=& \left( 1-\frac{f}{f_0^{(1)}}  \right) \Delta \Phi_1 \Theta(f-f_0^{(1)}) \nonumber\\
&& + \left( 1-\frac{f}{f_0^{(2)}}  \right) \Delta \Phi_2 \Theta(f-f_0^{(2)}),
\label{eq:phaseshift}
\end{eqnarray}
where $\Theta(x)$ denotes the usual step function. Specializing to r-mode excitations, 
the shifts $\Delta \Phi_i$, $i = 1, 2$ take the form \cite{Ma:2020oni}
\begin{equation}
\Delta \Phi_i 
= -2\frac{5\pi^2}{192} \left(\frac{4}{3}\right)^{2/3} 
\frac{\Omega^{2/3}_i}{\mathcal{M}^{10/3}}\mathcal{I}_i.
\end{equation}
Here $\Omega_i$, $i = 1, 2$ are the angular rotation frequencies of the neutron stars, and
$\mathcal{M} = (m_1 m_2)^{3/5}/(m_1 + m_2)^{1/5}$ is the chirp mass associated with 
the component masses $m_1$, $m_2$. The r-mode couplings $\mathcal{I}_i$ take the 
form 
\begin{equation}
\mathcal I_i
 = (\bar{I}^r_i)^2 \,m_i^4\, \text{sin}^2(\psi_i)\,
 \text{cos}^4(\psi_i/2)\,\left( 1 - \frac{m_i}{M}\right),
\label{eq:overlaps} 
\end{equation}
where $M = m_1 + m_2$ is the total mass, and $\psi_i$ are the angles between 
the neutron stars' spin vectors $\mathbf{S}_i$ and the orbital angular momentum. 
The r-mode \emph{overlaps} $\bar{I}^r_i$ depend on the neutron stars' internal 
structure.

Let us now discuss in turn our treatment of r-mode frequencies and overlaps.

\subsection{Universal relation for the r-mode frequencies}

For a slowly rotating neutron stars and in the Newtonian limit, the r-mode frequencies in 
the co-rotating frame are proportional to the angular rotation frequency $\Omega$, and 
given by \cite{Flanagan:2006sb}
\begin{equation}
  \omega_{\ell m} = -\frac{2m}{\ell(\ell+1)} \Omega.
\end{equation}
For given values of the spherical harmonic indices $(\ell, m)$, the corresponding 
resonance frequency in the inertial frame is
\begin{equation}
  \label{eqn:resonance}
  \omega_0 = m\Omega + \omega_{\ell m} = m\Omega - \frac{2m}{\ell(\ell+1)} \Omega.
\end{equation}
The associated GW frequencies appearing in Eq.~(\ref{eq:phaseshift}) are then $f_0 = 2\omega_0/(2\pi)$, 
or 
\begin{equation}
f_0 = \frac{1}{\pi} (m - \kappa)\,\Omega,
\label{eq:f0}
\end{equation}
with $\kappa = 2m/\ell (\ell + 1)$ in the Newtonian limit. 
For simplicity, in this work we will assume barotropic neutron stars, so that $|m| = l$, and
focus on $(\ell, m) = (2,2)$; within the Newtonian framework one then has $\kappa = 2/3$. 


As shown in \cite{Idrisy:2014qca}, in a relativistic treatment of slowly rotating stars, 
the resonance frequency can differ significantly from the Newtonian value. In that work, numerical 
values for $\kappa$ were computed for relativistic neutron stars described by a variety of 
tabulated EOSs, and a fitting formula was obtained for the function $\kappa(C)$, with 
$C = m/R$ the compactness, where $R$ is a neutron star's radius. However, in this work we 
are interested in the imprint of resonant r-modes on GW emission, where it is more convenient 
to have $\kappa$ as function of the dimensionless neutron star tidal deformability $\Lambda$, since
this is a parameter that enters directly into the waveform. 

\begin{figure}
  \centering
  \includegraphics[keepaspectratio, width=8.5cm]{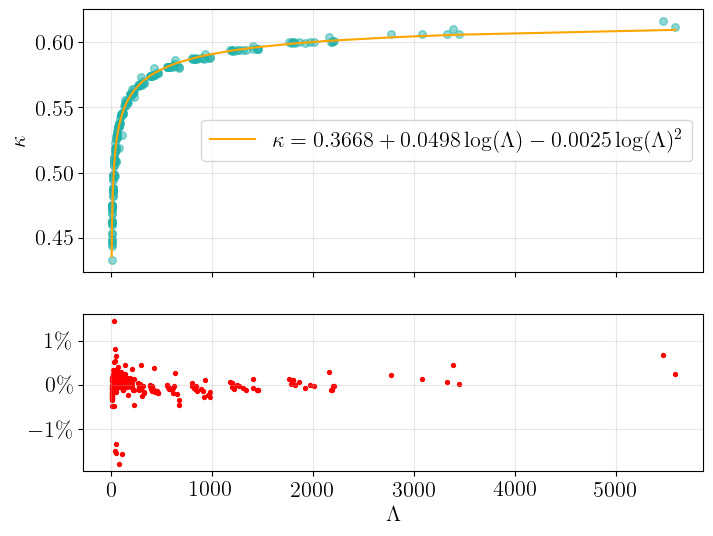}
  \caption[width = 0.8\textwidth]{Top panel: values of $\kappa$ and corresponding tidal deformabilities 
  $\Lambda$ computed for a variety of EOSs and compactnesses listed in \cite{Idrisy:2014qca} (green dots), 
  and the fit for $\kappa(\Lambda)$ of Eq.~\ref{eqn:kappafit}. Bottom panel: fitting residuals.}
  \label{fig:kappafit}
\end{figure}

With this in mind, for each value of $\kappa$ reported in \cite{Idrisy:2014qca} (Table II), 
we computed $\Lambda$ for each EOS and compactness listed. This was done using the TOV solver
available in LALSuite \cite{lalsuite}. As seen in Fig.~\ref{fig:kappafit}, there is a clear
dependence of $\kappa$ on the logarithm of $\Lambda$. Using a least-squares fitting method, 
we find that the functional dependence is approximated well by the universal relation
\begin{equation}
  \label{eqn:kappafit}
  \kappa = 0.3668 + 0.0498 \log(\Lambda) - 0.0025 \log^2(\Lambda),
\end{equation}
with fitting residuals at the $\mathcal{O}(1\%)$ level.

\subsection{Universal relation for the r-mode overlap}

Next we turn to the r-mode overlap. In the Newtonian approximation one has \cite{Ma:2020oni}
\begin{equation}
\bar{I}^r = \sqrt{\frac{1}{m^5} \int_0^R \rho r^6 d r},
\label{eq:newtonianoverlap}
\end{equation}
where $\rho$ is the density of the neutron star, $m$ its mass, and $R$ its radius. 
In \cite{Ma:2020oni}, a universal relation was obtained for the Newtonian $\bar{I}^r$ as function 
of $\Lambda$. In this work we instead follow the relativistic description of \cite{Gupta:2020lnv}, where 
the following expression was found:
\begin{equation}
\label{eq:IcalB}
 \bar{I}^r = \sqrt{\frac{15}{4 \pi} ( \Sigma_\text{stat} - \Sigma_\text{irr} )}.
\end{equation}
Here $\Sigma_\text{stat}$ and $\Sigma_\text{irr}$ are dimensionless static and 
irrotational magnetic Love numbers, respectively. Universal relations for these quantities
as function of $\Lambda$ were derived in \cite{JimenezForteza:2018rwr}, which for completeness
we display again here:
\begin{eqnarray} 
&&\log( -\Sigma_\text{irr} ) \nonumber\\
&&= -2.03 + 0.487\log(\Lambda) + 9.69\times10^{-3}\log^2(\Lambda) \nonumber\\
&&\,\,\,\,\,\,+ 1.03\times10^{-3}\log^3(\Lambda) - 9.37\times10^{-5}\log^4(\Lambda) \nonumber\\
&&\,\,\,\,\,\,+ 2.24\times10^{-6}\log^5(\Lambda), \\
&&\log(\Sigma_\text{stat}) \nonumber\\
&&= -2.66 + 0.786\log(\Lambda) -1.00\times10^{-2}\log^2(\Lambda) \nonumber \\
&&\,\,\,\,\,\,+ 1.28\times10^{-3}\log^3(\Lambda) - 6.37\times10^{-5}\log^4(\Lambda) \nonumber\\
&&\,\,\,\,\,\,+ 1.18\times10^{-6}\log^5(\Lambda).
\end{eqnarray}
Together with Eq.~(\ref{eq:IcalB}), these yield a relation for $\bar{I}^r(\Lambda)$. 
           
\section{Setup of the analyses}
\label{sec:setup}

In this section we describe the waveform model used, the Fisher matrix formalism for 
obtaining a multivariate Gaussian approximation of likelihoods, the parameterization 
used for reconstructing EOSs, and our framework for performing such reconstructions 
using multiple GW detections.

\subsection{Waveform model}
\label{subsec:waveform}

We will focus on the inspiral, and use an analytic, frequency domain waveform following
the stationary phase approximation \cite{Sathyaprakash:1991mt}, which takes the general form
\begin{align}
   \tilde{h}(f) = \mathcal{A} f^{-7/6}e^{i\Psi(f)}.
\end{align}
Here $\mathcal{A}$ collects parameters appearing in the amplitude: masses, sky position, 
orientation of the orbital plane, and luminosity distance to the source. The phase 
$\Psi(f)$ can be written as
\begin{equation}
\Psi(f) = \Psi_{\rm PP}(f) + \Psi_{\rm SO}(f) + \Psi_{\rm T}(f) + \Psi_r(f),
\end{equation}
where $\Psi_{\rm PP}$ has point particle contributions up to 3.5 post-Newtonian (PN) order, 
and $\Psi_{\rm SO}$ contains spin-orbit effects at 1.5PN; we consider spin-spin effects to 
be negligible in the case of binary neutron stars. (For a review of the post-Newtonian 
approximation, see \cite{Blanchet:2002av}.) $\Psi_{\rm T}$ contains tidal effects
at 5PN and 6PN orders \cite{Vines:2011ud}. Finally, $\Psi_r$ is as given in Eq.~(\ref{eq:phaseshift}), 
and we note that to good approximation, the angles $\psi_{1,2}$ entering Eq.~(\ref{eq:overlaps})
can usually be considered approximately constant even in the presence of spin precession
\cite{Kidder:1992fr,Kidder:1995zr,PhysRevD.49.6274}.

Without the r-mode contribution, the phase depends on the 8 free parameters
\begin{align}
\label{eq:no_rmode_params}
  \vec\theta_{nr} = ({t_c, \phi_c, m_1, m_2, \tilde {\Lambda}, \delta \tilde {\Lambda}, \chi_{sz}, 
  \chi_{az}}),
\end{align}
whereas with r-modes included, the free parameters are
\begin{align}
\label{eq:rmode_params}
  \vec\theta_{r} = ({t_c, \phi_c, m_1, m_2, \tilde {\Lambda}, \delta \tilde {\Lambda}, \Omega_1, 
  \Omega_2, \psi_1, \psi_2}).
\end{align}
Here $t_c$ and $\phi_c$ are respectively the time and phase of coalescence; $m_i$, $i = 1,2$ 
the component masses; $\chi_{sz} = (\chi_{1z} + \chi_{2z})/2$ and $\chi_{az} = (\chi_{1z} - \chi_{2z})/2$
respectively the symmetric and antisymmetric dimensionless spins, with $\chi_{1z}$ and $\chi_{2z}$ the 
spin components in the direction of orbital angular momentum; 
$\Omega_i$ the spin angular frequencies of the neutron stars; and $\psi_i$ the angles
between the dimensionless spins and the orbital angular momentum. We note that in the 
detector frame, the component masses $m_{1,2}^{\rm det}$ that directly enter the waveform are redshifted with 
respect to the source frame masses $m_{1,2}$: one has $m_{1,2}^{\rm det} = (1+z)\,m_{1,2}$, 
with $z$ the redshift. With a 
global network of third-generation (3G) observatories, for BNS signals the 
luminosity distance can be measured to $\mathcal{O}(1\%)$ accuracy for the loudest sources 
\cite{Zhao:2017cbb}, and with a cosmological model (e.g.~from Planck \cite{Ade:2013zuv}), distance 
can be converted to redshift. We will assume that other 3G observatories will be operational at the 
same time as ET, so that for the high-SNR sources considered below, the distance uncertainties will 
be sufficiently small that uncertainties on redshift can be neglected in the conversion between source 
frame and detector frame masses. Finally, $\tilde {\Lambda}$ and $\delta \tilde {\Lambda}$ are related to 
the individual neutron stars' tidal deformability parameters $(\Lambda_1, \Lambda_2)$ through 
\cite{Wade:2014vqa}
\begin{eqnarray} 
\tilde \Lambda &=& 
\frac{8}{3}\left[(1+7\eta-31\eta^2)(\Lambda_1+\Lambda_2)\right. \nonumber\\
&& \,\,\,\,\,\,\,+ \left. \sqrt{1-4\eta}(1+9\eta - 11\eta^2)(\Lambda_1-\Lambda_2)\right], \\
\delta \tilde \Lambda &=& \frac{1}{2}\left[\sqrt{1-4\eta}(1-\frac{13272}{1319}\eta 
+ \frac{8944}{1319}\eta^2)(\Lambda_1+\Lambda_2) \right. \nonumber\\ 
&& \,\,\,\,\,\,\,+ \left.(1-\frac{15910}{1319}\eta
+\frac{32850}{1319}\eta^2+\frac{3380}{1319}\eta^3)(\Lambda_1-\Lambda_2)\right], \nonumber\\
\end{eqnarray}
where $\eta= m_1m_2/(m_1+m_2)^2$ is the symmetric mass ratio. 

When comparing the cases with and without r-modes, we can use that 
\begin{equation}
\label{eq:chi_to_omega}
  \chi_{iz} = \bar{I}_i \Omega_i m_i \text{cos}(\psi_i)
\end{equation}
for $i = 1, 2$. Here $\bar{I}_i$ are normalized moments of inertia, for which one has 
the universal relation \cite{Yagi:2013bca}
\begin{eqnarray} 
\log\bar{I} &=& 1.47 + 8.17 \times10^{-2}\log(\Lambda) + 1.49\times10^{-2}\log^2(\Lambda)  \nonumber \\
&& +  2.87\times10^{-4}\log^3(\Lambda) - 3.64\times10^{-5}\log^4(\Lambda). \nonumber\\ 
\end{eqnarray}

\subsection{Spectral parametrization of the EOS}

For the EOS we will use the so-called spectral parameterization in terms of the adiabatic 
index $\Gamma(p)$, defined as \cite{Koliogiannis:2018hoh,Lindblom:2010bb}
\begin{equation}
\label{eq:adiabatic_index}
\Gamma(p) = \frac{\epsilon+p}{p}\frac{dp}{d\epsilon},
\end{equation}
where $\epsilon$ is energy density and $p$ is pressure. 
The EOS $\epsilon(p)$ is obtained from the adiabatic index by writing the above equation as
\begin{equation}
\frac{d\epsilon}{dp} = \frac{\epsilon(p)+p}{p\Gamma(p)},
\end{equation}
or 
\begin{equation}
\epsilon(p) =\frac{\epsilon_0}{\mu(p)}+\frac{1}{\mu(p)}\int_{p_0}^{p} \frac{\mu(p')}{\Gamma(p')} dp', 
\label{eq:ep}
\end{equation}
where 
\begin{equation}
\mu(p) =\exp \left(-\int_{p_0}^{p} \frac{1}{p'\Gamma(p')} dp'\right),
\label{eq:mu}
\end{equation}
with $\epsilon_0=\epsilon(p_0)$ a constant of integration. The 
EOS $\epsilon(p)$ can in principle be solved for arbitrary adiabatic index $ \Gamma(p)$, 
but here we will spectrally decompose it in terms of a set of polynomial basis functions,
\begin{equation}
\Gamma(p) = \exp\bigg(\sum_{k=0}^n{\gamma_k x^k}\bigg),
\label{eq:spectral}
\end{equation}
where $x=\log(p/p_0)$ is a dimensionless pressure variable, 
$p_0$ is some reference pressure, and values of $n$ up to 3 tend to allow for  
accurate representations of a variety of 
EOSs \cite{Lindblom:2010bb}. 

For given coefficients $\gamma_k$ in Eq.~(\ref{eq:spectral}), the TOV solver available in 
LALSuite \cite{lalsuite} numerically performs the integrals in Eqs.~(\ref{eq:mu}) 
and (\ref{eq:ep}) to obtain $\epsilon(p)$ and rest mass density $\rho(p)$, 
which is inverted to arrive at $p(\rho)$. The reference pressure $p_0$, also called
the stitching pressure, is fixed to $5.3716\times10^{32}$ dyne cm${^{-2}}$; below 
this pressure, the EOS called SLY \cite{Douchin:2001sv} is stitched on. The coefficients $\gamma_k$, $k = 0, \ldots, 3$
are given uniform priors with ranges $\gamma_0 \in [0.2, 2.0]$, 
$\gamma_1 \in [-1.6, 1.7]$, $\gamma_2 \in [-0.6, 0.6]$, and $\gamma_3\in [-0.02, 0.02]$. 
The speed of sound $v_s = \sqrt{dp/d\epsilon}$ is 
restricted to $v_s < 1.1\,c$, where the 10\% leeway is to allow for imperfect parameterization. 
Finally, the adiabatic index is confined to $\Gamma(p) \in [0.6, 4.5]$ \cite{LIGOScientific:2018cki}.


\subsection{Analysis framework}

Let $\theta^a$ be the components of the parameter vector $\vec\theta$, which in our 
case will be either the one from Eq.~(\ref{eq:no_rmode_params}) or (\ref{eq:rmode_params}), 
depending on whether or not r-modes are taken into account.  
For GW events with sufficiently high SNR and assuming stationary, Gaussian noise, the likelihood 
function for the signal parameters $\theta^a$ will approximately take the form of a 
multivariate Gaussian centered on the true values $\hat{\theta}^a$ \cite{Finn:1992xs,PhysRevD.46.5236}. 
Defining $\Delta\theta^a = \theta^a - \hat{\theta}^a$, the likelihood becomes
\begin{equation}
\label{eq:likelihood}
L(\Delta\theta^a) = \mathcal{N}e^{-\frac{1}{2}\Gamma_{ab}\Delta \theta^a \Delta \theta^b},
\end{equation}
where $\mathcal{N}$ is a normalization factor, and sums over $a$ and $b$ are implied. The
\emph{Fisher matrix} is given by
\begin{equation}
\label{eq:fisher}
      \Gamma_{ab} = \left\langle \frac{\partial h}{\partial \theta^a} \bigg| 
      \frac{\partial h}{\partial \theta^b} \right\rangle,
\end{equation}
where the noise-weighted inner product $\langle\,\cdot\,|\,\cdot\,\rangle$ is defined as
\begin{equation}
\label{eq:innerproduct}
      \langle h|g \rangle = 4 \Re \int_{f_{\text{low}}}^{f_{\text{high}}} \frac{\tilde{h}^* (f) 
      \tilde{g}(f)}{S_n(f)} df,
\end{equation}
with $S_n(f)$ the one-sided noise power spectral density (PSD).

Strictly speaking, Eq.~(\ref{eq:likelihood}) with (\ref{eq:fisher}) pertains to 
signals $\tilde{h}(f)$ as seen in a \emph{single} detector, while the baseline design for Einstein 
Telescope assumes three V-shaped detectors arranged in a triangle \cite{Punturo:2010zza,Hild:2010id}.
In principle one would then have to project the gravitational wave polarizations $\tilde{h}_+(f)$, 
$\tilde{h}_\times(f)$ onto each of the three detectors using the appropriate antenna pattern functions, 
construct a separate Fisher matrix for each detector, and take the sum of these to obtain the final
Fisher matrix. In this work we will consider sources from a catalog constructed as in 
\cite{Samajdar:2021egv}, with SNRs computed for a triangular ET. However, since here 
we will mainly be interested in information coming from the phasing, for our purposes 
it will suffice to compute a single Fisher matrix with the ET-D PSD \cite{Hild:2010id} for 
an L-shaped detector, and with the SNRs set to the values obtained from the triangular ET. 
In Eq.~(\ref{eq:innerproduct}) we take the lower frequency cut-off $f_{\rm low}$ to be 5 Hz. 
For $f_{\rm high}$ we choose the nominal frequency of the innermosts stable circular orbit (ISCO): 
$f_{\rm high} = 1/(6^{3/2}\pi (1+z) M)$. We note that for stiff EOSs, the two neutron stars
might touch before ISCO is reached \cite{Agathos:2015uaa}, but the effect of this
for the purposes of EOS measurements is marginal \cite{Wade:2014vqa}.

Since we are interested in how well one can measure the EOS with BNS inspiral signals, 
we need to choose a ``true'' EOS, which we take to be FPS \cite{Friedman:1981qw}, with 
$(\gamma_0, \gamma_1, \gamma_2, \gamma_3) = (1.1561, -0.0468, 0.0081, -0.0010)$. Given 
a choice of masses $m_1$, $m_2$, we let the true values of the tidal 
deformabilities be $\Lambda_1 = \Lambda_{\rm FPS}(m_1)$, $\Lambda_2 = \Lambda_{\rm FPS}(m_2)$, 
where $\Lambda_{\rm FPS}(m)$ is the dependence set by the given EOS.


The main question in this work is how well ET will be able to determine the EOS, i.e., 
with what accuracy the EOS parameters 
$\vec{E} \equiv (\gamma_0, \gamma_1, \gamma_2, \gamma_3)$ will be measured. To this end, 
consider the posterior probability density function $p(\vec{E}, \vec{\theta}'|d, \mathcal{H}, \mathcal{I})$,
where $\vec\theta'$ denotes the parameters in (\ref{eq:no_rmode_params}) or (\ref{eq:rmode_params}) 
\emph{except} for $\tilde{\Lambda}$ and $\delta\tilde{\Lambda}$, which can be calculated from 
$\vec{E}$ together with $m_1$, $m_2$; other than that, $d$ denotes the data for a given 
signal, $\mathcal{H}$ a waveform model, and $\mathcal{I}$ any background information we may possess.
Bayes' theorem tells us that 
\begin{equation}
p(\vec{E}, \vec{\theta'}|d, \mathcal{H}, \mathcal{I})
= \frac{p(d|\vec{E}, \vec{\theta}', \mathcal{H}, \mathcal{I})\,p(\vec{E}, \vec{\theta}'|\mathcal{H}, \mathcal{I})}{p(d|\mathcal{H}, \mathcal{I})},
\label{eq:Elikelihood}
\end{equation}
where $p(\vec{E}, \vec{\theta}'|\mathcal{H}, \mathcal{I})$ is the prior probability 
density for $\vec{E}$ and $\vec\theta'$, $p(d|\vec{E}, \theta', \mathcal{H}, \mathcal{I})$
the likelihood, and $p(d|\mathcal{H}, \mathcal{I})$ the evidence, which is set by the 
requirement that the posterior probability density be normalized. Clearly 
$p(d|\vec{E}, \vec{\theta}', \mathcal{H}, \mathcal{I})$ is not quite the same as the likelihood 
$L(\Delta\theta^a)$ obtained from the Fisher matrix in Eq.~(\ref{eq:likelihood}), 
but given true values $\hat{\theta}^a$ for the parameters entering the waveform, 
it is possible to relate the two by writing 
\begin{eqnarray}
L(\Delta\theta^a) &=& p(d|\vec{\theta}, \mathcal{H}, \mathcal{I}) \nonumber\\
&=& p(d|\{\tilde\Lambda_{\vec{E}}(m_1, m_2), \delta\tilde\Lambda_{\vec{E}}(m_1, m_2), 
\vec{\theta}'\}, \mathcal{H}, \mathcal{I}) \nn\\
&=& p(d|\{\vec{E}, \vec{\theta}'\}, \mathcal{H}, \mathcal{I}),
\label{eq:identification}
\end{eqnarray}
where in the second line, $\tilde\Lambda_{\vec{E}}(m_1, m_2)$ and $\delta\tilde\Lambda_{\vec{E}}(m_1, m_2)$
are the $\tilde\Lambda$ and $\delta\tilde\Lambda$ obtained from the component masses for 
an EOS with the given parameters $\vec{E}$, and the likelihood in the last line is the 
one appearing in Eq.~(\ref{eq:Elikelihood}). 

The way we will proceed is then as follows. Using the Fisher matrix we compute the likelihood of 
Eq.~(\ref{eq:likelihood}), and with the identification of Eq.~(\ref{eq:identification})
this is turned into a likelihood in terms of the EOS parameters $\vec{E}$ and waveform 
parameters $\vec\theta'$. We choose flat priors for all of the individual parameters, and using the 
\texttt{emcee} sampler \cite{Foreman_Mackey_2013} we obtain the posterior density 
in Eq.~(\ref{eq:Elikelihood}). By integrating out the $\vec\theta'$, this gives us a posterior
density for the $\vec{E}$, $p(\vec{E}|d, \mathcal{H}, \mathcal{I})$.\footnote{In principle we could have 
set $\tilde\Lambda = \tilde\Lambda_{\vec{E}}(m_1, m_2)$ and 
$\delta\tilde\Lambda = \delta\tilde\Lambda_{\vec{E}}(m_1, m_2)$ directly in the Fisher matrix, 
and obtained 1-sigma uncertainties on the components of $\vec{E}$ from the covariance matrix in 
the usual way \cite{Finn:1992xs,PhysRevD.46.5236}. However, introducing too many additional parameters 
can lead to ill-conditioned Fisher matrices, which will cause problems with numerical inversion.} 
So far we have focused on a single signal, but given a catalog of detected signals 
$d_1, d_2, \ldots, d_N$ that are considered independent,\footnote{If all neutron stars
follow the same EOS (as is implicitly assumed here), then in reality the $d_1, d_2, \ldots, d_N$
will not be completely independent. Thus, our way of obtaining combined results may be 
somewhat sub-optimal, but this just means that our conclusions about ET's ability to reconstruct the 
EOS will be on the conservative side.} information from all of them can be combined to obtain
\cite{DelPozzo:2013ala}
\begin{eqnarray}
&&p(\vec{E}|d_1, d_2, \ldots, d_N, \mathcal{H}, \mathcal{I}) \nonumber\\
&&= p(\vec{E} | \mathcal{H}, \mathcal{I})^{1-N} \prod_{n=1}^N p(\vec{E}|d_n, \mathcal{H}, \mathcal{I}),
\label{eq:combinedposterior}
\end{eqnarray}
where in practice the posteriors in the product are Gaussian kernel density estimates 
of the ones obtained directly with \texttt{emcee}. Through the TOV solver of LALSuite
\cite{lalsuite}, the above combined posterior distribution for $\vec{E}$ leads to 
a distribution over equations of state $p(\rho)$.

Sources are picked from a catalog constructed as in \cite{Samajdar:2021egv}. 
For binary neutron stars this assumes uniformly distributed source frame component masses, where for 
the primary mass $m_1 \in [1\,M_\odot, M_{\rm max}]$ and for the secondary mass $m_2  
\in [1\,M_\odot, m_1]$; here we take $M_{\rm max}$ to be the maximum mass supported by our 
reference EOS, which is $M_{\rm  max} = 2.03\,M_\odot$. As explained in Sec.~\ref{subsec:waveform}, 
we will assume that for the highest-SNR sources, distances can be measured with 
sufficient accuracy that uncertainties on redshift can be neglected in the conversion between source 
frame masses and detector frame masses. The sources were distributed over comoving distance 
according to a particular prediction for the 
merger rate as a function of redshift; for details we refer to the original paper \cite{Samajdar:2021egv}. 
ET is likely to see tens of thousands of BNSs per year, but it is reasonable to expect that 
most of the information will come from the loudest sources. Hence we consider the 
20 loudest sources in the catalog, which have SNRs between 154 and 368. 
The angles between the spin vectors and the orbital angular momentum are taken to be 
uniform on the sphere. The rotation angular frequencies of the neutron stars 
are taken to be uniform in the intervals $\Omega_i \in [0, 2\pi\times 45]$ Hz, $i = 1, 2$ 
\cite{Lorimer:2008se}. Given these ranges for the $\Omega_i$ and the assumed
detector lower frequency cut-off of $f_{\rm low} = 5$ Hz, there is a small chance for the r-mode 
GW frequency $f_0$ in Eq.~(\ref{eq:f0}) to be below $f_{\rm low}$ for one or both neutron stars
in a binary, but this will not be the case for any of the simulated sources considered here. We will 
perform our analyses with r-modes included, in which 
case the free parameters are the ones in Eq.~(\ref{eq:rmode_params}), and without r-modes, 
in which case the free parameters are the ones in (\ref{eq:no_rmode_params}), with the values 
for $\chi_{1z}$ and $\chi_{2z}$ set according to Eq.~(\ref{eq:chi_to_omega}).


\section{Results}
\label{sec:results}

We are now ready to assess the accuracy with which ET will be able to reconstruct the EOS 
of dense nuclear matter. However, in order to check the validity of our method, 
we first compare results for a simulated GW170817-like source seen in Advanced LIGO with those 
that were obtained in reality for GW170817. Then we turn to ET and perform an analysis on the 20 
loud sources mentioned above, with or without r-modes included.

\subsection{Analysis of a GW170817-like signal}

Let us consider a signal with properties similar to that of GW170817, the binary neutron 
star signal discovered with LIGO and Virgo in 2017 \cite{LIGOScientific:2017vwq}. Looking 
at maximum-likelihood parameter values obtained in \cite{LIGOScientific:2018hze}, 
we take $(m_1, m_2) = (1.44, 1.27)\,M_\odot$, $\chi_{1z} = 0.022$, and $\chi_{2z} = 0.0081$. 
The SNR is set to 32.4, and for the Fisher matrix we take the PSD to be the one 
for LIGO Livingston at the time of the detection, setting $f_{\rm low} = 23$ Hz and
$f_{\rm high} = 1625.36$ Hz. The effects of r-modes are not included, since they would 
have had no impact in this case. Given that we do not know what is the true equation of state, 
we consider the same EOS as in the rest of this paper (namely FPS), which for the given masses 
yields $(\Lambda_1, \Lambda_2) = (387.0, 872.2)$. 

\begin{figure}
    \centering
   \includegraphics[keepaspectratio, width=9.5cm]{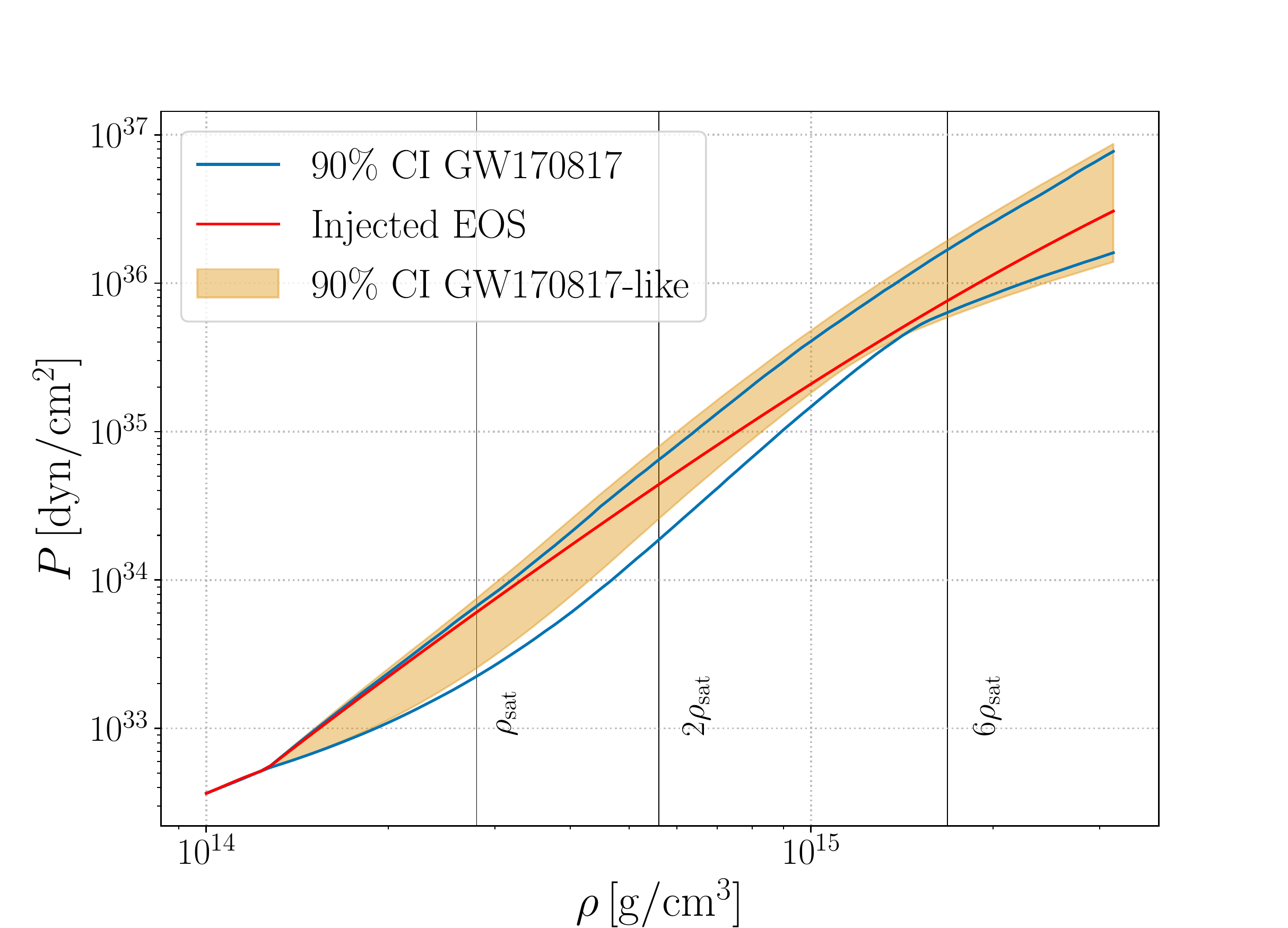}
    \caption[width = 0.8\textwidth]{Shown is pressure versus density for a GW170817-like signal 
    whose EOS is FPS (the red curve), with a 90\% credible region (orange). This is compared with 
    the 90\% credible region that was actually obtained for GW170817 (black curves)
    \cite{LIGOScientific:2018cki}. 
    Vertical lines indicate a few multiples of the nuclear saturation density, $\rho_{\rm sat}$.
    }  
    \label{fig:GW170817like}
\end{figure}

With the formalism described in the previous section, the above parameters lead to a ``measurement'' 
of $p(\rho)$. Fig.~\ref{fig:GW170817like} shows the underlying $p(\rho)$ together with 
a 90\% credible region. This is compared with the 90\% credible region for $p(\rho)$ that was 
actually obtained for this event \cite{LIGOScientific:2018cki}. The qualitative similarity 
lends confidence to our methodology.

\subsection{EOS reconstruction with Einstein Telescope}

\begin{figure}
    \centering
   \includegraphics[keepaspectratio, width=9.5cm]{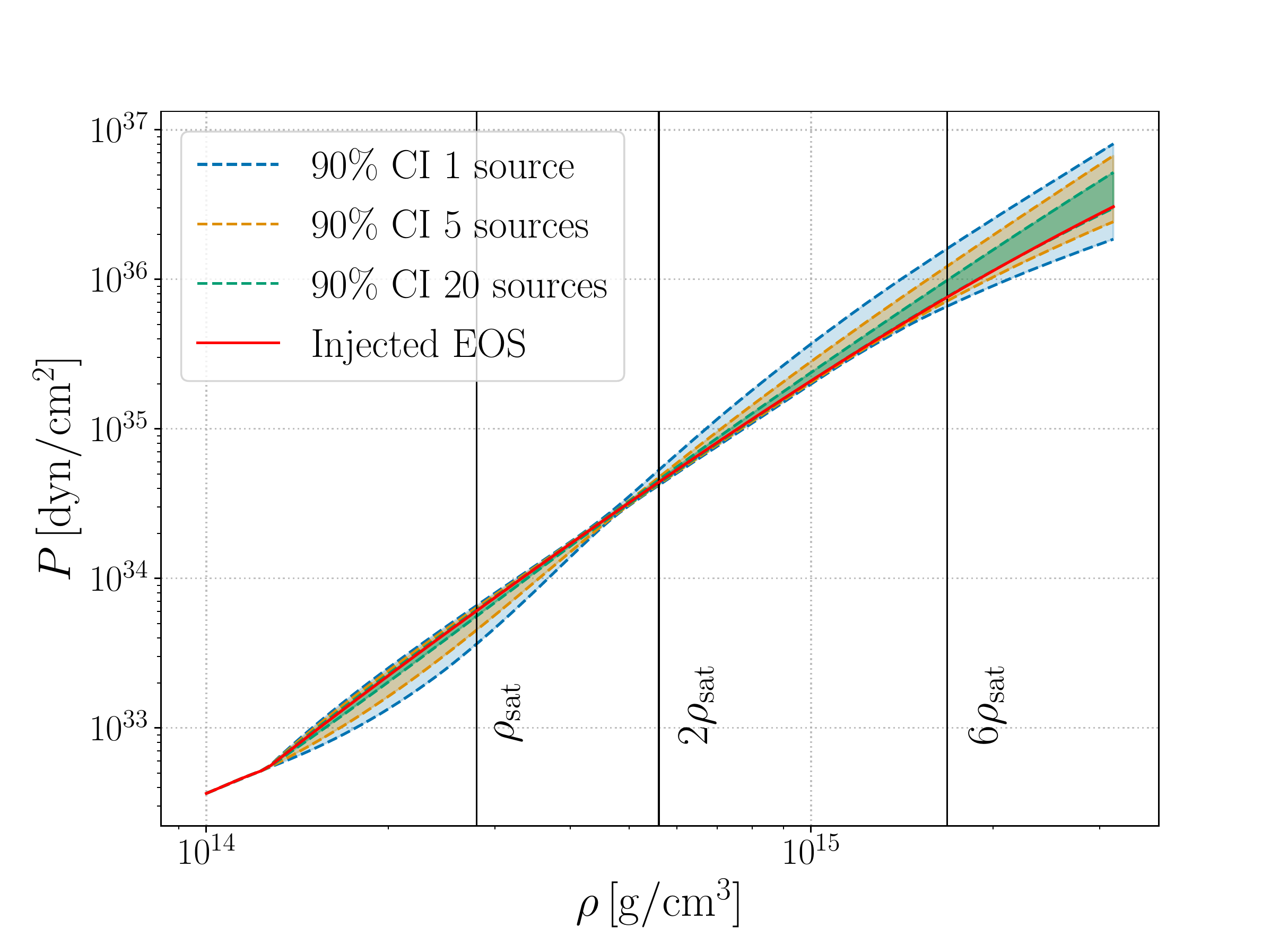}
    \caption[width = 0.8\textwidth]{90\% credible regions for pressure versus density from the loudest source, the 
    5 loudest sources, and the 20 loudest sources in ET, in the case where r-modes are 
    included. 
    }  
    \label{fig:1_5_20_rmodes}
\end{figure}

Next we turn to ET. Fig.~\ref{fig:1_5_20_rmodes} shows the EOS recovery with only the loudest source, 
the 5 loudest sources, and the 20 loudest sources in our simulated catalog, 
with r-modes included in the way that was explained in the previous section.   
An improvement in measurement accuracy with increasing number of sources is clearly in evidence. 
We also note the narrowing of the 90\% credible region near twice the nuclear saturation density, 
$\rho = 2\rho_{\rm sat}$, this being the approximate average density for most of the neutron stars 
in our BNSs, given the EOS we picked. At $2\rho_{\rm sat}$ and for 20 sources, the width of the 90\% 
credible interval for the pressure is $4.77 \times 10^{33}\,\mbox{dyn}/\mbox{cm}^2$, to be compared with 
$8.33 \times 10^{34}\,\mbox{dyn}/\mbox{cm}^2$ for our simulated GW170817 as seen
by Advanced LIGO. At other densities, ET improves somewhat less on the results
for GW170817; the widths $\Delta P_{90\%}(\rho)$ of the 90\% credible intervals for pressure at a few different values 
for density are shown in Table \ref{tab:widths}, for different numbers of sources in ET, with and 
without r-modes. Note how at $2\rho_{\rm sat}$, a single loud source in ET improves the pressure estimation by a factor 
of $\sim 3$ over GW170817, but when combining information from 20 sources the improvement is by a factor 
of $\sim 17$ with r-modes included (and a factor of $\sim 11$ without r-modes). Still 
at $2\rho_{\rm sat}$, the gain from r-modes reaches $\sim 50\%$.

\begin{widetext}
\begin{center}
\begin{table}
\begin{tabular}{ |l|c|c|c| }
\hline
Sources & $\Delta P_{90\%}(\rho_{\rm sat})\,\left[\mbox{dyn}/\mbox{cm}^2\right]$ & $\Delta P_{90\%}(2\rho_{\rm sat})\,\left[\mbox{dyn}/\mbox{cm}^2\right]$ & $\Delta P_{90\%}(6\rho_{\rm sat})\,\left[\mbox{dyn}/\mbox{cm}^2\right]$  \\
\hline
\hline
GW170817-like  & $7.08 \times 10^{33}$ & $8.33 \times 10^{34}$ & $2.07 \times 10^{36}$ \\
ET, 1 source   & $3.47 \times 10^{33}$ ($3.67 \times 10^{33}$) & $2.62 \times 10^{34}$ ($3.33 \times 10^{34}$) & $1.50 \times 10^{36}$ ($1.85 \times 10^{36}$) \\
ET, 5 sources  & $2.16 \times 10^{33}$ ($2.97 \times 10^{33}$) & $1.16 \times 10^{34}$ ($1.79 \times 10^{34}$) & $8.59 \times 10^{35}$ ($1.11 \times 10^{36}$) \\
ET, 10 sources & $8.50 \times 10^{32}$ ($1.71 \times 10^{33}$) & $5.94 \times 10^{33}$ ($9.37 \times 10^{33}$) & $5.14 \times 10^{35}$ ($7.33 \times 10^{35}$) \\
ET, 20 sources & $6.41 \times 10^{32}$ ($1.29 \times 10^{33}$) & $4.81 \times 10^{33}$ ($7.54 \times 10^{33}$) & $3.92 \times 10^{35}$ ($5.77 \times 10^{35}$) \\
\hline
\end{tabular}
\caption{The widths of the 90\% credible intervals for the pressure at different densities, for our 
GW170817-like analysis, and for ET with different numbers of sources. In the case of ET, the numbers in 
brackets are without r-modes.}
\label{tab:widths}
\end{table}
\end{center}
\end{widetext}


\begin{figure}
    \centering
   \includegraphics[keepaspectratio, width=9.5cm]{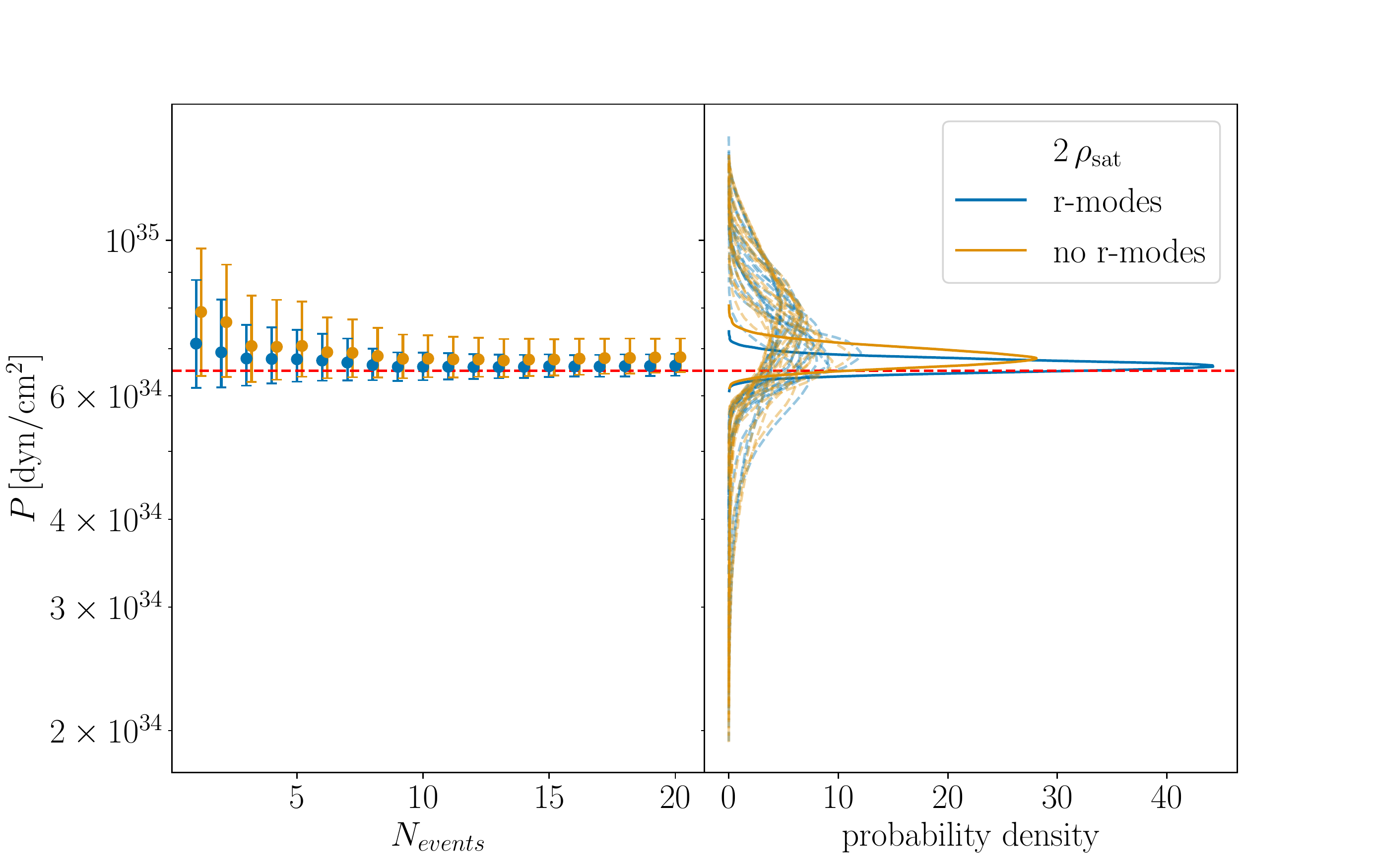}
    \caption[width = 0.8\textwidth]{Left panel: The improvement in the measurement of pressure 
    at twice the nuclear saturation density; shown is the evolution of 90\% credible intervals when 
    going from 1 source to 20 sources, where the blue includes r-modes while the orange does not. 
    From left to right, sources are being added in order of decreasing SNR. 
    Right: The individual and combined probability distributions for pressure.
    }  
    \label{fig:2nsat}
\end{figure}

Thus, more advanced GW observatories will dramatically improve our knowledge of the EOS, 
both through increased sensitivity and by seeing a larger number of sources. The largest improvement 
happens near densities that actually occur in neutron stars.  
In Fig.~\ref{fig:2nsat} we show the accuracy 
in the measurement of pressure at $\rho = 2\rho_{\rm sat}$, with and without r-modes. 
The benefit of seeing r-modes is clearly in evidence. Note how in both cases, most of 
the measurement accuracy comes from combining information from the $\sim 10$ loudest sources.

\section{Summary and conclusions}
\label{sec:conclusions}

We have investigated how well one will be able to determine the equation of state of neutron 
stars with Einstein Telescope, given that this observatory will detect tens of BNS 
inspirals per year for which the SNR will be in the hundreds. In doing so we 
have taken into account the effect of resonant r-modes, which provides additional information 
about the EOS. The latter were treated fully relativistically, both in terms of the 
resonance frequency and the r-mode overlap; for the former we introduced a new 
universal relation linking it to the neutron star tidal deformability, and for the 
latter we utilized the recent treatment in \cite{Gupta:2020lnv}. A reference EOS was chosen, 
and general EOSs were represented in terms of the so-called spectral parameterization.
Simulations were performed based on the Fisher approximation of the likelihood; when 
identifying the tidal deformabilities $\Lambda_i$ with the ones obtained from an EOS 
with \emph{a priori} unknown parameters $\vec{E}$, sampling the likelihood leads to a PDF
for $\vec{E}$. From this one can reconstruct the EOS in terms of pressure as a function of density. We tested our formalism on a simulated 
GW170817-like source, and found the accuracy of our $p(\rho)$ reconstruction to be similar 
to what was obtained in reality. We then turned to ET, and focused on the 20 loudest sources 
in a simulated ``catalog'' with realistic assumptions for the merger rate as a function of
redshift.

The estimates arrived at in this paper are necessarily somewhat crude, due to the approximations 
made, and especially the limitations of the Fisher matrix formalism. We only aimed to 
give a rough sense of how accurately ET would be able to pin down the EOS; however, the 
comparison we made for GW170817 between results from our formalism and actual EOS measurements 
lends confidence to the reliability of our predictions.   

As expected, ET will dramatically improve our knowledge of the EOS compared with what was
gleaned from GW170817. In terms of $p(\rho)$, the main improvement comes at densities around twice the 
nuclear saturation density, since this is the typical average density of a neutron star. At that density, 
ET with 20 loud sources will improve pressure measurements by a factor $\sim 17$ over GW170817. 
We found that most of the information will come from the $\sim 10$ loudest sources.

We also saw that the inclusion of r-modes leads to an improvement in our ability to measure 
the EOS, at the 50\% level in terms of pressure at twice saturation density. This may seem 
small compared to what is suggested by \cite{Ma:2020oni} for a single source, but 
in that work, an optimistic scenario was assumed where the source had an extremely large 
SNR (of $\sim 1500$), and values for the angles $\psi_1$, $\psi_2$ between the spins and the orbital 
angular momentum were chosen so as to nearly maximize the effect of r-modes on the gravitational 
waveform. By contrast, our sources had SNRs between 154 and 368, and randomly chosen angles $\psi_i$. 

In this work we only considered information about the EOS coming from the inspiral part 
of the signal. However, ET will also have access to the merger and post-merger \cite{Faber:2012rw}, 
which will lead to further improvements in the measurement of the EOS. Assessing the effect 
of the latter is left for future work.

\begin{acknowledgments} 

P.K.G., A.P., P.T.H.P., G.K., J.J., and C.V.D.B.~are supported by the research programme 
of the Netherlands Organisation for Scientific Research (NWO). We are grateful to 
Tim Dietrich, Tanja Hinderer, and Jan Steinhoff for useful discussions at an early 
stage of the work. This research has made use of data, software and/or web tools 
obtained from the Gravitational Wave Open Science Center (https://www.gw-openscience.org), a 
service of LIGO Laboratory, the LIGO Scientific Collaboration and the Virgo Collaboration. 
LIGO is funded by the U.S. National Science Foundation. Virgo is funded by the French
Centre National de Recherche Scientifique (CNRS), the 
Italian Istituto Nazionale della Fisica Nucleare (INFN) 
and the Dutch Nikhef, with contributions by Polish and Hungarian institutes.

\end{acknowledgments}

\bibliography{refs}

\onecolumngrid

\end{document}